\documentclass[12pt,preprint,round,colon]{emulateapj}
\begin{document}

\shortauthors{Metchev et al.}
\shorttitle{AO Imaging of AU Mic}

\title{Adaptive Optics Imaging of the AU Microscopii Circumstellar Disk: 
Evidence for Dynamical Evolution}

\author{Stanimir A.\ Metchev, Joshua A.\ Eisner, Lynne A.\ Hillenbrand}
\affil{California Institute of Technology, Division of Physics, Mathematics 
\& Astronomy, MC 105--24, Pasadena, California 91125}
\email{metchev, jae, lah@astro.caltech.edu} 
\and
\author{Sebastian Wolf}
\affil{Max-Planck-Institut f\"{u}r Astronomie, K\"{o}nigstuhl 17, D--69117 
Heidelberg, Germany}
\email{swolf@mpia-hd.mpg.de}

\accepted{}

\begin{abstract}
We present an $H$-band image of the light scattered from circumstellar dust 
around the nearby (10~pc) young M star AU~Microscopii (AU~Mic,
GJ~803, HD~197481), obtained with the Keck adaptive optics system.
We resolve the disk both vertically and radially, 
tracing it over 17--60~AU from the star.
Our AU Mic observations thus offer the
possibility to probe at high spatial resolution
($0.04\arcsec$ or 0.4~AU per resolution element) for morphological 
signatures of the debris disk on Solar-System scales.  
Various sub-structures
(dust clumps and gaps) in the AU~Mic disk may point to the existence of
orbiting planets.
No planets are seen in our $H$-band image down to a limiting mass of
1~$M_{\rm Jup}$ at $>$20~AU, although the existence of smaller planets can not
be excluded from the current data.  
Modeling of the disk surface brightness distribution 
at $H$-band and $R$-band, in conjunction with the optical to sub-millimeter
spectral energy distribution, allows us to constrain the disk geometry 
and the dust grain properties.  We confirm the nearly edge-on orientation 
of the disk inferred from previous observations, and deduce an inner 
clearing radius $\leq$10~AU.
We find evidence for a lack of small grains in the inner ($<$60~AU)
disk, either as a result of primordial disk evolution, or because of
destruction by Poynting-Robertson and/or corpuscular drag.
A change in the power-law index of the surface brightness profile is observed
near 33~AU, similar to a feature known in the profile of the $\beta$~Pic 
circumstellar debris disk.  By comparing the time scales for
inter-particle collisions and Poynting-Robertson drag between the two systems,
we argue that the breaks are linked to one of these two processes.

\end{abstract}


\section{INTRODUCTION}

The existence of dust disks around main-sequence stars has been known
since the first days of the {\sl Infrared Astronomy Satellite (IRAS)} 
mission, when \citet{aumann_etal84} reported the detection of strong
far-infrared (far-IR) excess emission toward Vega ($\alpha$~Lyr).  Over 
200 other main-sequence stars have since been reported to possess such 
excesses, found almost exclusively with {\sl IRAS} and the {\sl Infrared 
Space Observatory} \citep[e.g.,][]{backman_paresce93, mannings_barlow98,
silverstone00, habing_etal01, spangler_etal01,
laureijs_etal02, decin_etal03}, though recently also with {\sl Spitzer}
\citep[e.g.,][]{meyer_etal04, gorlova_etal04}, and through ground-based
sub-millimeter observations \citep{carpenter_etal04}.  Too old to possess
remnant primordial dust, that would be cleared by radiation pressure and
Poynting-Robertson (P-R) drag within several million years (Myr) in the 
absence of
gas, these stars owe their far-IR excess to emission by ``debris
disks,'' formed by the collisional fragmentation of larger bodies
\citep[the so-called ``Vega phenomenon'';][and references 
therein]{backman_paresce93}.  Subsequent imaging at optical to 
millimeter wavelengths of the nearest sub-sample of Vega-like
stars has resolved intricate disk-like structures, with gaps and
concentrations \citep{holland_etal98, holland_etal03, greaves_etal98, 
schneider_etal99, krist_etal00, koerner_etal01, wilner_etal02, 
weinberger_etal02, clampin_etal03}.  

The most favored explanation for
such structures is the gravitational perturbation by 
embedded planets orbiting at semi-major axes comparable to the disk
size \citep{moromartin_malhotra02, kenyon_bromley04}.
The existence of perturbing planets may be revealed
by clumps of dust trapped in mean motion resonances, as has been 
suggested for 
Vega \citep{wilner_etal02, wyatt03}, $\epsilon$~Eri \citep{ozernoy_etal00,
quillen_thorndike02}, and Fomalhaut
\citep[$\alpha$~PsA;][]{wyatt_dent02, holland_etal03}, and 
observed by the {\sl Cosmic Background Explorer} satellite along the 
Earth's orbit
\citep{reach_etal95}.  Stochastic collisions between large planetesimals
that result in dust clumps
lasting several hundreds of orbital periods are another means
of producing disk asymmetries \citep{stern96}.
Spiral density waves \citep[as seen in the disk
of HD~141569A, and inferred around $\beta$~Pic;][]{clampin_etal03,
kalas_etal00}, and warps in the disk inclination \citep[as in the disk
of $\beta$~Pic;][]{heap_etal00, wahhaj_etal03}, may indicate
perturbation by nearby stars \citep{kalas_etal00, kalas_etal01, 
kenyon_bromley02a, augereau_papaloizou04}.  Finally, dust migration in 
a gas-rich disk can produce
azimuthally symmetric structures, as observed in the HR~4796A 
circumstellar disk \citep{takeuchi_artymowicz01}.

High-resolution imaging observations,
such as those of HR~4796A \citep{schneider_etal99}, $\beta$~Pic
\citep{heap_etal00}, HD~141569A \citep{weinberger_etal99} and TW~Hya 
\citep{krist_etal00, weinberger_etal02} with the 
{\sl Hubble Space Telescope (HST)} 
can help single out the most likely physical process behind the
disk morphology.   The resolution achievable with adaptive optics in the
near-infrared on large ground-based telescopes rivals that of {\sl HST}, and
is the method employed in this paper for investigating disk structure.

The young \citep[8--20~Myr;][]{barrado_y_navascues_etal99,
zuckerman_etal01b} M1~V \citep{keenan83} star 
AU~Mic has a known 60$\micron$ excess from {\sl IRAS}, likely due to
orbiting dust \citep{song_etal02}.  Because of its relative proximity
\citep[{\sl Hipparcos} distance of 9.94$\pm0.13$~pc;][]{perryman_etal97}, 
AU~Mic 
is a good target for high-resolution imaging of scattered light to
characterize the circumstellar disk morphology.
Recent 450$\micron$ and 850$\micron$ observations by \citet{liu_etal04} 
confirmed the existence of dust, and follow-up $R$-band (0.65$\micron$)
coronagraphic imaging revealed a nearly edge-on 
disk extending 210~AU \citep{kalas_etal04} from the star.  
Because the age of AU~Mic is larger than
the collision timescale between particles in the disk (0.5--5~Myr at 200~AU),
\citet{kalas_etal04} infer that most of the dust particles have
undergone at least one (destructive) collision, and hence the AU~Mic
disk is a debris disk.  However, because the P-R time
scale for 0.1--10$\micron$ particles at $\gtrsim$100~AU from the star is 
greater than the stellar age, \citeauthor{kalas_etal04} expect that most 
of the disk at large radii consists of primordial material.
\citeauthor{liu_etal04} find a fractional infrared luminosity,
$L_{IR}/L_\ast = 6\times10^{-4}$ and fit
the far-IR to sub-millimeter excess by a 40~K modified
blackbody with constant emissivity for 
$\lambda<100\micron$ and following $\lambda^{-0.8}$ for longer
wavelengths.  From the lack of excess at 25$\micron$,
\citeauthor{liu_etal04} infer an inner disk edge at 17~AU from the star,
or 1.7$\arcsec$ at the distance of AU~Mic.  They speculate that such a 
gap may have been opened by an orbiting planet, that, given the youth of
the system, could be
detectable in deep adaptive optics (AO) observations in the near IR.  
Such high contrast observations could also be
used to search for signatures of planet/disk interaction.

The AU~Mic circumstellar disk is not resolved with the 14$\arcsec$ beam of the
JCMT/SCUBA observations of \citeauthor{liu_etal04}, and the 
\citeauthor{kalas_etal04} optical coronagraphic observations are insensitive
to the disk at separations $<$5$\arcsec$ because of the large sizes of their
occulting spots (diameters of 6.5$\arcsec$ and 9.5$\arcsec$), and because 
of point-spread 
function (PSF) artifacts.  Taking advantage of the higher angular
resolution and dynamic range achievable with adaptive optics on
large telescopes, \citet{liu04} used the Keck AO system to investigate 
the disk morphology at separations as small as 15--20~AU from the star.
We present our own set of Keck AO data that confirms \citeauthor{liu04}'s
observations, and places upper limits on the presence of
potential planetary companions.
In addition, we combine our spatially resolved $H$-band 
information with the $R$-band imaging data from \citet{kalas_etal04}, and with
the optical to sub-millimeter spectral energy distribution (SED) of AU~Mic 
from \citet{liu_etal04}, to put self-consistent
constraints on the disk morphology and the dust properties, as done
previously for $\beta$~Pic \citep*{artymowicz_etal89}.
We use a full three-dimensional radiative transfer code to model
simultaneously the SED and the $H$-band and $R$-band surface brightness
profiles (SBPs).  We find that our model consisting of a single dust
population does not reproduce the observed break in the $H$-band SBP, 
whereas a
two-component dust model, as proposed for $\beta$~Pic, fits the data well.  
Drawing from a comparison with the $\beta$~Pic system, we deduce that dynamical
evolution of the disk provides the simplest explanation for the morphology of
the SBPs of both disks.

\section{OBSERVATIONS AND DATA REDUCTION \label{sec_obs}}

We observed AU~Mic at 
$H$ band (1.63$\micron$) with the NIRC2 instrument
(Matthews et al., in prep.) and the AO system \citep{wizinowich_etal00}
on the Keck~II telescope.  The data were acquired on 5 
June, 2004 under photometric conditions.
We employed coronagraphic spots of different
sizes (0.6$\arcsec$--2.0$\arcsec$ diameter) to block out the light from
the star.  The observations were taken with the wide, 0.04$\arcsec$~pix$^{-1}$ 
camera in NIRC2, which delivers a $41\arcsec\times41\arcsec$ field of view 
(FOV) on the 1024$\times$1024 InSb Alladin-3 array.

We obtained nine 54~sec exposures at $H$ band, three
with each of the 0.6$\arcsec$, 1.0$\arcsec$, and 2.0$\arcsec$-diameter
coronagraphic spots.  We observed the nearby
(2.4$\degr$ separation) M2/3~III star HD~195720 as a PSF standard, 
with similar colors, but 0.9~mag brighter than AU~Mic at
$H$.  We spent equal amounts of time on the target, on
the PSF star, and on sky.
The observations were carried out according to the following sequence,
repeated 3 times: 3 exposures of AU~Mic, 
3 exposures of HD~195720, and 3 exposures of the blank sky (taken at
three dithered positions: 60$\arcsec$, 50$\arcsec$, and 40$\arcsec$ away
from the PSF star).  The total on-source exposure time was 8.1~min.
Throughout the observations, the field rotator was set in ``position 
angle mode,'' preserving the orientation of the sky on the detector.  The image
quality was estimated from the 
Strehl ratios of point sources observed at higher
spatial resolution (with the narrow camera, 0.01$\arcsec$/pix) at the
beginning of each night.  Our $H$ band images had Strehls of 17--20\%.

\begin{figure}
\plotone{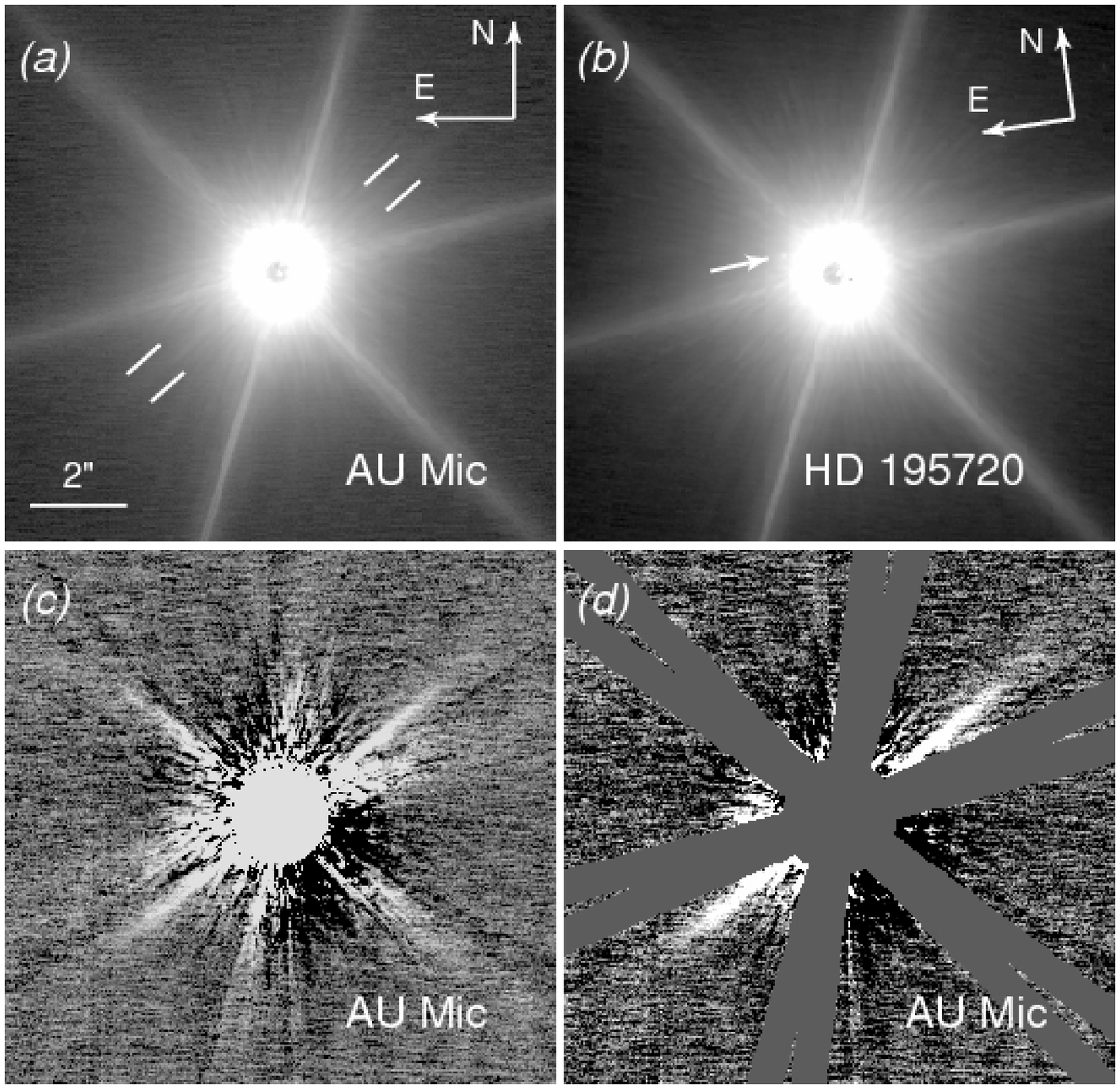}
\figcaption{$H$-band images of AU~Mic. {\it (a)} A median-combined image 
of three sky-subtracted 54.3~sec exposures with the 0.6$\arcsec$
coronagraph.  The star is visible through the semi-transparent
coronagraph.  The disk is discernible as a faint pair of diametrically
opposite spikes along the SE--NW direction (traced by the two pairs of
parallel lines).
{\it (b)} An image of the PSF star, HD~195720, with the same coronagraph, 
scaled to the intensity of the AU~Mic image in panel {\it (a)}.  An arrow
points to the location of a faint projected companion
(Section~\ref{sec_detlims}).
{\it (c)} The final median-combined image of AU~Mic obtained from
PSF-subtracted images with 0.6$\arcsec$, 1.0$\arcsec$ and 2.0$\arcsec$ spot
diameters.  All surface brightness photometry is performed on this image.
{\it (d)}.  Same as in {\it (c)}, but with an overlaid digital mask covering
the star and the telescope diffraction pattern.
The residual noise in the central region (3.4$\arcsec$-diameter circle) 
is greater than the surface brightness of the edge-on disk.
\label{fig_aumic}}
\end{figure}

\begin{figure}
\plotone{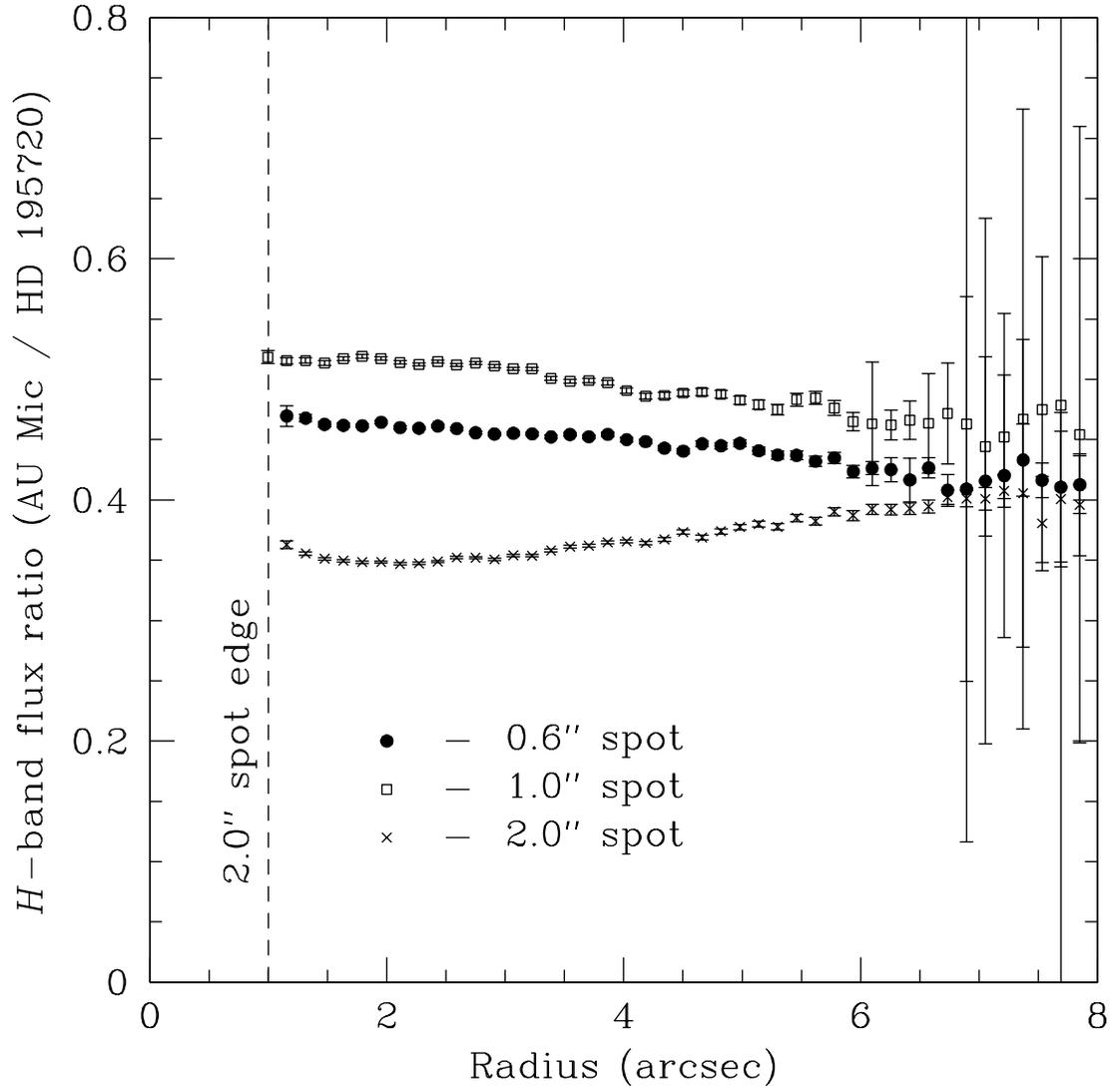}
\figcaption{Median ratios $f(r)$ of the radial profiles of AU~Mic and 
HD~195720 (the
PSF star) at $H$ band for the images taken with the various coronagraphs.
The vertical dashed line shows the edge of the largest coronagraph.  Our PSF
images were multiplied by $f(r)$ before subtracting them from the
corresponding AU~Mic images.  For any given coronagraphic spot size, $f(r)$
varies by less than 15\% in the range $r=1-8\arcsec$.
\label{fig_radprofs}}
\end{figure}

Data reduction followed the standard steps of sky-subtraction,
flat-fielding, and bad-pixel correction.  The images in the individual
sets of 3 exposures were then median-combined to improve the
sensitivity to faint objects.  The AU~Mic dust
disk was barely discernible at this stage (Figure~\ref{fig_aumic}a).  
To enhance the visibility of the dust disk, we subtracted the stellar
PSF. The PSF was obtained by first rotating the image of HD~195720
to match the orientation of the diffraction pattern in the AU~Mic image, 
and then scaling it by a centrally symmetric function $f(r)$ to match
the radial dependence of the AU~Mic profile.
The objective of the scaling was to compensate for variations in the seeing
halo between AU~Mic and the control star caused by changing atmospheric 
conditions and fluctuating quality of the AO correction.  The function 
$f(r)$ was obtained as the median
radial profile of the ratio of the AU~Mic to the HD~195720 images, with
the telescope spikes and the edge-on disk masked.  
Figure~\ref{fig_radprofs} shows $f(r)$ for the images with the three different
coronagraphic spot sizes.  Two remarks on this procedure should be made
here.  First, the function $f(r)$ does not vary by more than 
15\% for any given spot size, and for large radii tends to 0.4~--- the $H$-band
flux ratio between AU~Mic and the PSF star.  Second, because the AU~Mic
disk happens to be edge-on, the centrally symmetric scaling of the PSF
does not introduce any spurious features in the result, and
so does not interfere with the morphology of the disk.  The procedure
would not be viable for disks that are far from edge-on.

The scaled version of HD~195720 for
the image with the 0.6$\arcsec$ coronagraph is
presented in Figure~\ref{fig_aumic}b.  Panel (c) of
Figure~\ref{fig_aumic} shows the final AU~Mic image, obtained by
median-combining all coronagraphic exposures.  In panel (d)
a digital mask has been employed to enhance the appearance of the
circumstellar disk.  The mask encompasses the innermost 1.7$\arcsec$ from
AU~Mic, as well as the hexagonal diffraction spikes from 2 of the 3 image 
sequences where they did not subtract well (as seen in panel (c) of the
Figure).

For flux calibration we adopted the 2MASS magnitude for AU~Mic
($H=4.831\pm0.016$) and relied on the residual transmission of the
NIRC2 coronagraphic spots.  The PSF star, HD~195720, is
saturated in 2MASS, and therefore unusable for flux calibration.  
We measured the flux from AU~Mic through the
1.0$\arcsec$ and 2.0$\arcsec$ coronagraphs in a 6~pix (0.24$\arcsec$)
diameter aperture.  From non-coronagraphic images taken with the
10~mas/pix camera on NIRC2, we found that this aperture contained 68\% of
the total power in the PSF.  The $H$-band transmissivity of the 
2$\arcsec$ NIRC2
coronagraph was measured at $(7.32\pm0.24)\times10^{-4}$ (extinction of
$7.84\pm0.03$~mag).
AU~Mic is known to exhibit a large $V$-band photometric amplitude 
(0.35~mag) due to star spots with a period of
4.9~days \citep{torres_ferrazmello73}.  Although we do not expect measurable
variability over the $\approx$1~hour time span of our observations,
our absolute flux calibration is uncertain.  Nevertheless,
in the near IR the contrast between the spots and the stellar
photosphere is less pronounced than in the optical, so the uncertainty is 
also smaller: likely of order $\lesssim$0.1~mag.

The absolute orientation of the dust disk arms was calibrated through
observations of a binary star standard, WDS~18055+0230, with well-known 
orbital elements \citep[grade~1;][]{hartkopf_mason03b, pourbaix00}.
The $y$-axis of the NIRC2 detector was measured to
be offset by $1.24\degr\pm0.10\degr$ clockwise from North.  All position 
angles quoted below have been corrected for this offset.

\section{RESULTS \& ANALYSIS}

\subsection{Circumstellar Dust Morphology \label{sec_morph}}

The disk is seen out to a distance of $\sim$6$\arcsec$ (60~AU) from the 
star in our combined $H$-band image (Figure~\ref{fig_aumic}c,d).  
Inwards it can be traced inwards to $\approx$1.7$\arcsec$ (17~AU) from 
the star, at which point residual speckle noise from the PSF subtraction
overwhelms the emission from the disk.  Thus, our imaging data cannot directly
test the existence of the proposed disk clearing interior to 17~AU from 
the star \citep{liu_etal04}.  We confirm the sharp mid-plane
morphology of the disk \citep{kalas_etal04, liu04}, indicating a
nearly edge-on orientation, and resolve the disk thickness, with the SE arm
appearing somewhat thicker (FWHM~= 2.8--4.4~AU) than the NW arm (FWHM~=
2.2--4.0~AU).
There is also evidence of an increase in the FWHM of each of the arms with
separation: from 2.2--2.8~AU at 20~AU to 4.0--4.4~AU at 40~AU from the star,
indicating a potential non-zero opening angle of the disk.
Within 5$\arcsec$ of AU~Mic, the position angles (PAs)
of the two sides of the disk are nearly 180$\degr$ away from each other:
we measure PA~= 310.1$\degr\pm$0.2$\degr$ for the NW arm, and 
PA~=129.5$\degr\pm$0.4$\degr$ for the SE arm.  These PAs are in agreement 
with those reported in \citet[][311.4$\degr\pm$1.0$\degr$ and
129.3$\degr\pm$0.8$\degr$]{liu04}, though more accurate, likely as a
result of our proper calibration of the orientation of the NIRC2
detector (Section~2).

\begin{figure}
\plotone{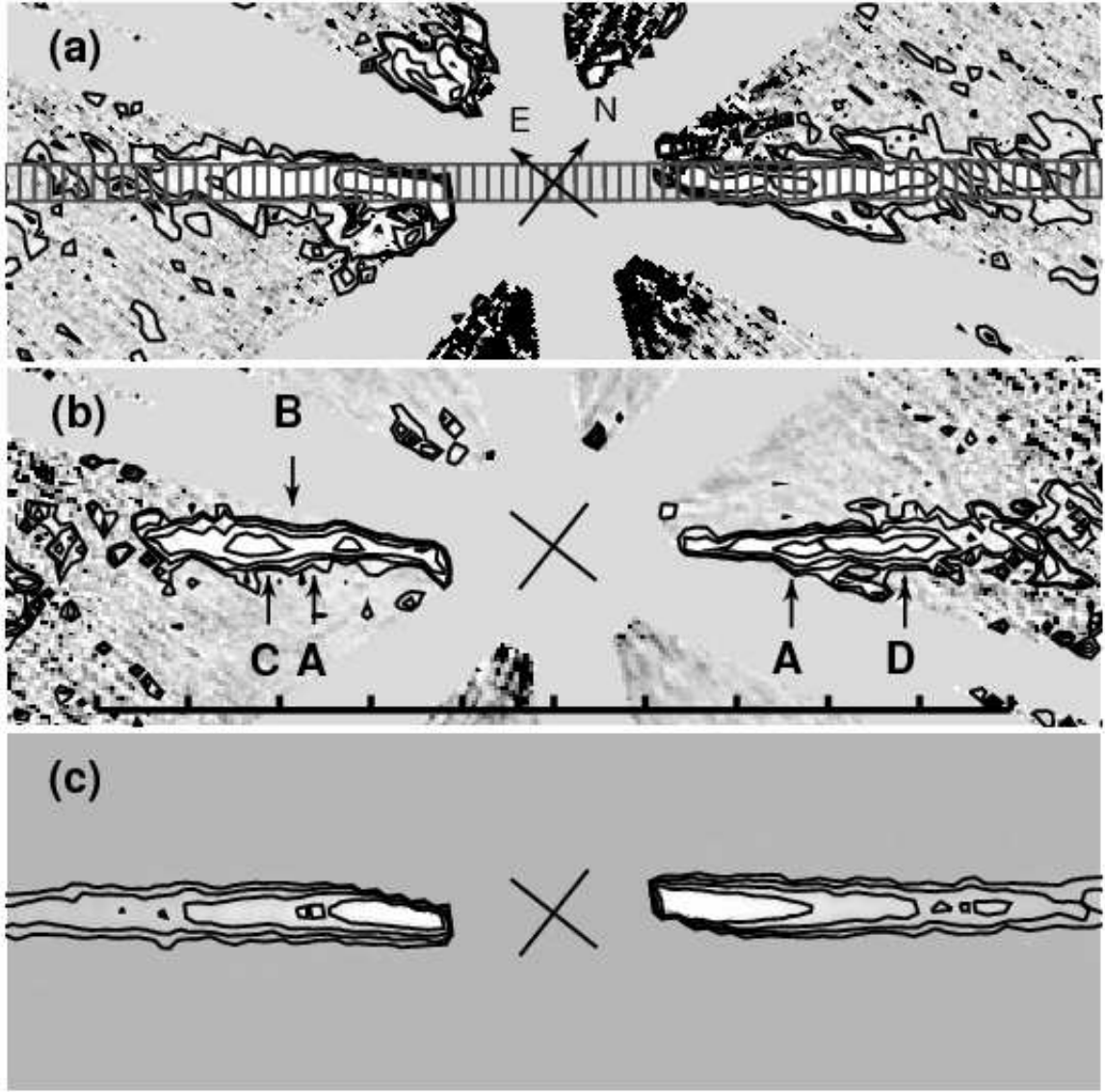}
\figcaption{$12.0\arcsec\times4.0\arcsec$ images of the AU~Mic disk, with the 
SE arm oriented horizontally.
(a) Locations of the photometry regions for measuring the disk
surface brightness superimposed on the final, masked, image of AU~Mic.  The
regions are $0.16\arcsec\times0.48\arcsec$, and sample
1--2 FWHMs of the disk thickness.  The circular mask is 2$\arcsec$ 
in radius.  The crossed arrows mark the location of AU~Mic.  The contour levels
trace surface brightness from 17.5~mag~arcsec$^{-2}$ to 
14.3~mag~arcsec$^{-2}$ in steps of 0.8~mag~arcsec$^{-2}$.
(b) Small-scale structure in the AU~Mic disk.  The capital letters 
correspond to sub-structures identified by \citet{liu04}.  The bar at 
the bottom is 10$\arcsec$ (100~AU) long, extending from $-50$~AU to 
+50~AU along the disk plane, with tick marks every 10~AU.  To enhance the
appearance of the clumps in the disk, we have multiplied the pixel values
by the square of the distance from the star.  The contour levels follow a
squared intensity scale.  
(c) The preferred scattered light model of the AU~Mic disk at $H$-band,
created using the MC3D code
(Section~4;
Table~\ref{tab_model}).  The same software mask
as in the other two panels has been applied.  The contour levels follow the 
same spacing as in panel (a).
No background noise is added, though Poisson-noise ``clumps'' due to
low signal-to-noise of the model can be seen.  These do not represent discrete
physical structures.
\label{fig_contours}}
\end{figure}

The radial SBP of the disk was measured on the reduced 
image (Figure~\ref{fig_aumic}c; before applying the digital mask)
using the {\sc IRAF} task {\sc polyphot}.
The photometry regions are indicated by the rectangles overlapped 
onto a contour map of the image in Figure~\ref{fig_contours}a.  
We used
4~pix~$\times$ 12~pix ($0.16\arcsec\times0.48\arcsec$) rectangular 
apertures, where the long side of the rectangular regions was chosen to span 
1--2 FWHMs of the disk thickness, and was aligned normally to the disk arm.
The distance between the aperture centers was 4~pix ($0.16\arcsec$).
Even though the image was PSF- and sky-subtracted, to offset for flux
biases introduced by the centrally symmetric scaling of the PSF
(Section~2),
we employed additional background subtraction,
with the background flux estimated as
the median pixel value in 0.16$\arcsec$-wide concentric annuli centered on 
the star.  The
photometric uncertainty was estimated as the quadrature sum of the standard
deviation of the background and the photon noise from the disk.
For the standard deviation of the background we adopted the root mean
square (r.m.s.) of the pixel values in the annulus, multiplied by 
$\sqrt{\pi/2}$ \citep[to properly account for the standard deviation of the
median;][]{kendall_stuart77}, and normalized 
by the size of the photometry aperture.  

\begin{figure}
\plotone{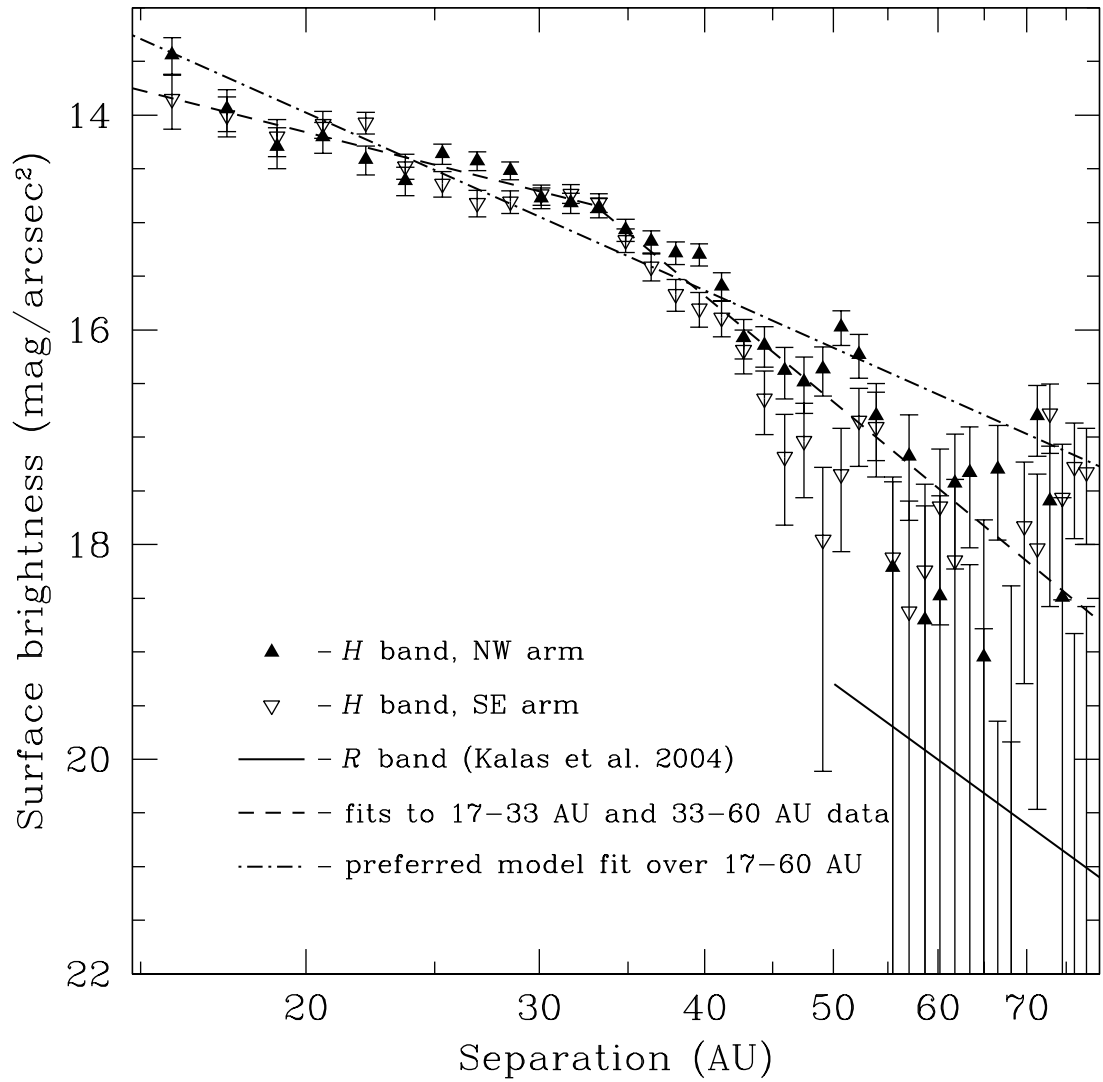}
\figcaption{$H$-band surface brightness profiles of the NW (upward
pointing solid triangles) and the SE (downward pointing open triangles)
arms of the AU~Mic disk.  
A gradual flattening of the SBPs of both arms is observed inwards of
30--40~AU.  The dashed lines represent the power-law fits to the mean SBP at
17--33~AU (index of $-1.2\pm0.2$), and 33--60~AU (index of $-4.0\pm0.6$).
The solid line represents the mean $R$-band SBP from
\citet{kalas_etal04} with a power-law index of $-3.75$.
The dot-dashed line shows are preferred model with a power-law index $-2.2$,
matching that ($-2.3\pm0.2$) of the mean SBP over 17--60~AU.
\label{fig_sbps}}
\end{figure}

The radial SBPs of the NW (upward pointing triangles) and SE (downward 
pointing triangles) arms of the projected disk are shown in 
Figure~\ref{fig_sbps}.  The two SBPs agree well throughout
the region over which we can trace the disk (17--60~AU).
Unlike the observed shape of the $R$-band SBP at 50--210~AU,
the $H$-band SBP at 17--60~AU from the star cannot be fit
by a single power law.  Instead, the SBPs of both the NW and the SE arms 
``kink'' and flatten inwards of 30--40~AU, with the transition being
more abrupt in the SE arm at $\approx$33~AU, and more gradual in the NW
arm.  While the power-law exponent of the mean SBP
over the entire range (17--60~AU) is $-2.3\pm0.2$, over 17--33~AU 
separations it flattens to $-1.2\pm0.3$, while over 33--60~AU, it increases to 
$-4.0\pm0.6$.
These are consistent with the measurements of \citet{liu04} over the same 
separation range.  

A closer look at the SBP of the NW and SE arms reveals several
small-scale asymmetries, all of which can be linked to regions of
non-uniform brightness in the AU~Mic disk (Figure~\ref{fig_contours}a). 
The sub-structure is enhanced by scaling the reduced image by a
radially symmetric function centered on the star, with magnitude 
proportional to the radius squared (Figure~\ref{fig_contours}b).  
The lettered structures denote features identified in the (deeper) image 
of \citet{liu04}: clumps of enhanced
emission (A, and C), a gap (B), and a region elevated with respect to the inner 
disk mid-plane (D).  In general, we confirm the presence of these features in 
the AU~Mic
disk, although the gap (B) and the clump (A) appear misplaced by $\sim$5~AU
toward the star in our image.  The NW arm also looks more uniform in
brightness between 17--40~AU in our image compared to that in \citet[][Figures
3 and 4]{liu04}, where clump A is very prominent.  These discrepancies
may be caused by residual speckle noise from the PSF subtraction in
either data set.
In addition to the features described by \citeauthor{liu04}, we see a faint
clump in
the NW arm at $\gtrsim$50~AU from the star, coincident with the location of
the bump in the SBP of this arm (Figure~\ref{fig_sbps}).  
The concentration is not reported by \citeauthor{liu04},
and being at a relatively low signal-to-noise ($\approx$3),
may be a noise spike.

\subsection{Disk Luminosity, Optical Depth and Geometry \label{sec_tau}}

The integrated disk brightness (over 17--60~AU from the star) is 
2.7$\pm$0.8~mJy at $H$ band; hence $L_{\rm scat} / L_{\ast} = 
2.3\times10^{-4}$.  This is comparable to the fractional dust luminosity
in emitted mid-IR to sub-mm light, $f_d=L_{IR}/L_\ast =6\times10^{-4}$
\citet{liu_etal04}, and hence suggests that the disk mid-plane may be
optically thin to radiation at wavelengths as short as $\sim$1.5$\micron$,
at the peak of the AU~Mic spectrum.  Indeed, $f_d$ is
similar to that of other known debris disks
\citep[$10^{-5}-10^{-3}$; e.g.,][]{sylvester_mannings00, habing_etal01,
spangler_etal01}, all of which are optically thin to ultra-violet 
and optical light in the direction perpendicular to the disk plane 
($\tau_\perp \ll 1$).  In the mid-plane, the optical depth of grains
along a radial line from the star to infinity is $\tau_\parallel \sim f_d/\sin
\delta$ if the grains are in a ``wedge'' or ``flaring'' disk with thickness
proportional to radius and opening angle $2\delta$ \citep{backman_paresce93}.
Because of the generally unknown viewing geometry of circumstellar disks, 
$\tau_\parallel$ tends to be poorly constrained.
Assuming edge-on orientation ($i=90\degr$), we can estimate the
maximum allowed opening angle $2\delta$ from the observed disk thickness.  For
smaller values of $i$, $\delta$ will be smaller because of projection
effects.  Assuming a perfectly flat, thin disk, we find a
lower limit on the inclination $i>87\degr$ over 20--50~AU.  The projected
appearance of an inclined disk of zero thickness
would however be inconsistent with the apparent thickening of the disk with
increasing separation (Section~\ref{sec_morph}).  Therefore, the disk likely
has non-zero scale height and/or opening angle, and is viewed within only a
degree of edge-on.  To put a limit on $\delta$, we observe that
at a radius of 40~AU the disk FWHM is $\sim$4~AU.  Thus we obtain that 
for $i\lesssim90\degr$, $\delta\lesssim3\degr$.  
Hence, $\tau_\parallel \geq 6\times10^{-3}$, and 
probably less than unity, i.e., the disk is optically thin in the radial 
direction.  

\subsection{Detection Limits on Sub-Stellar Companions
\label{sec_detlims}}

\begin{figure}
\plotone{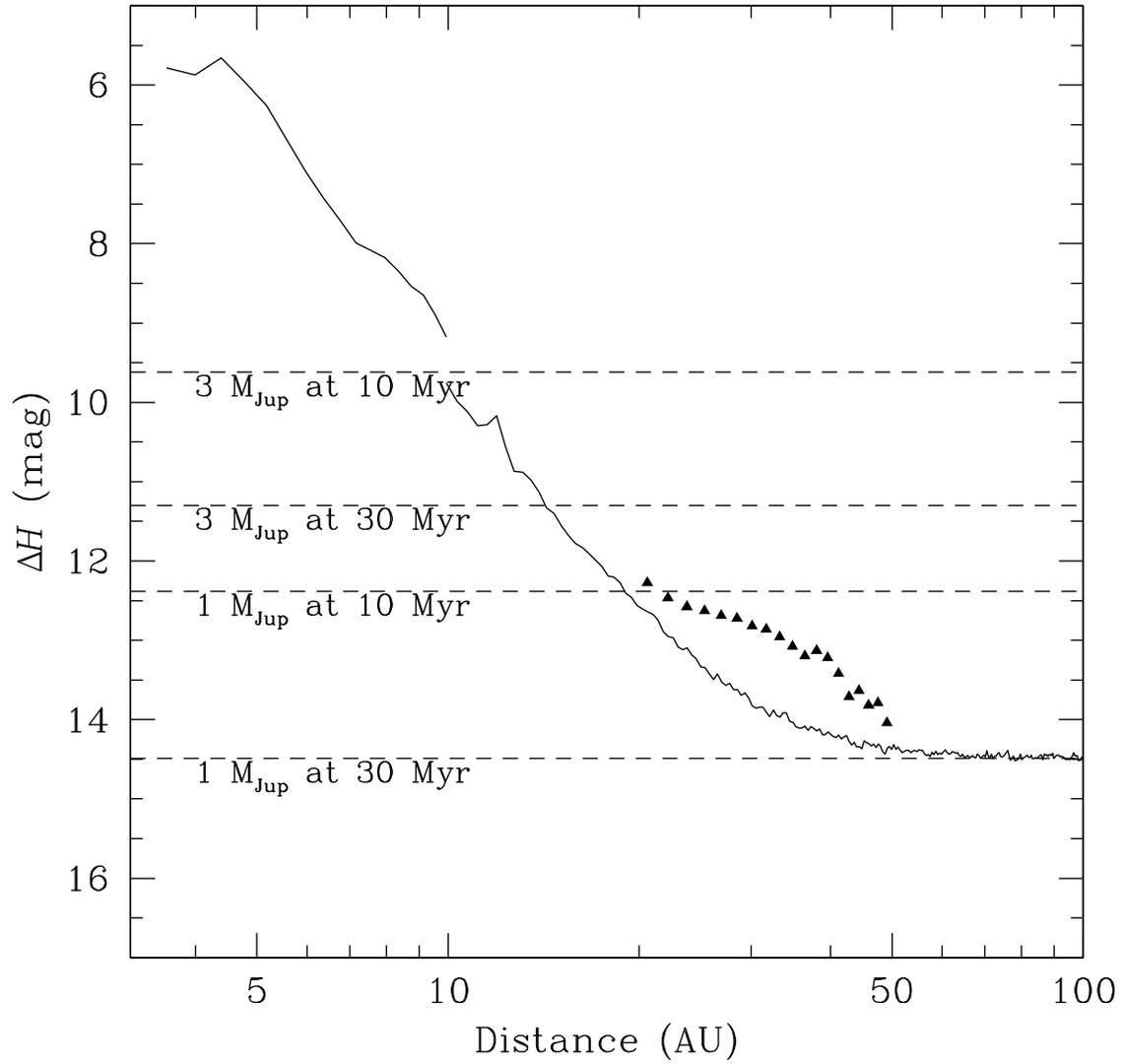}
\figcaption{$H$-band 5$\sigma$ detection limits for companions to AU~Mic.  The solid
line delineates the limits in regions away from the disk.  The break at
10~AU ($1\arcsec$) corresponds to the edge of the largest 
(2$\arcsec$-diameter) coronagraphic
spot.  To determine the detection limits at separations $\leq$1$\arcsec$, 
we used only the series of images taken with the 0.6$\arcsec$ spot,
constituting a third of the total exposure time.  Thus, our 
sensitivity at small separations is somewhat worse than what the
extrapolation from distances $>$10~AU would predict.  The triangle symbols
trace the poorer sensitivity to point sources in the plane of the disk.  
Limited experimentation with planting artificial sources in the image
confirmed these detection limits.
The dashed lines indicate the expected contrast for 1 and 3$M_{\rm Jup}$
planets around AU~Mic \citep{burrows_etal97} for system ages of 10 and
30~Myr.  
\label{fig_detlims}}
\end{figure}

Dynamical influence by embedded planets is a frequently invoked
explanation for substructure in dust disks.  Because of its youth,
proximity, and late spectral type, AU~Mic is an ideal target for direct
imaging of planets.  However,
no point sources are seen in our combined 8.1~min PSF-subtracted
$H$ band exposure.  Figure~\ref{fig_detlims} delineates our
5$\sigma$ sensitivity limits as a function of angular separation from 
the star in the PSF-subtracted image.  The 1$\sigma$ level at each 
distance is defined as the r.m.s.\ scatter of the pixel 
values in one-pixel wide annuli centered on the star.  This was divided
by $\sqrt{28.3-1}$ to adjust
for the finite size of the aperture used for photometry (six-pixel
diameter, or an area of 28.3~pix$^2$).  The sensitivity to point sources
within the disk is up to 1~mag poorer because of the higher photon noise.
This is shown in Figure~\ref{fig_detlims} with the points, each of
which has the photon noise from the disk signal added in quadrature.
These detection limits, calculated in a statistical manner, were
confirmed through limited experiments with artificially planted stars.

At the location of the inferred gap in the SE
arm ($\sim$25--30~AU), we can detect planets down to 1 Jupiter
mass ($M_{\rm Jup}$) for an assumed stellar age of 10~Myr and using 
brown-dwarf cooling models from \citet{burrows_etal97}.  Dynamical models of 
planet-disk interactions
in other systems exhibiting similar disk morphology ($\epsilon$~Eri,
$\alpha$~Lyr) require planets 0.1--3~$M_{\rm Jup}$ 
\citep{quillen_thorndike02, wilner_etal02}.
Provided that the clumps in the AU~Mic disk are caused by such a
planet, our point source detection limits constrain its mass to the
lower part of this range.

A faint candidate companion is seen around our PSF star,
HD~195720 ($H=3.88\pm0.24$ from 2MASS).  The 
object is 9.5$\pm$0.2~mag fainter at $H$, at a projected 
separation of 1.19$\arcsec$ and PA of 81$\degr$ (Figure~\ref{fig_aumic}b).  
Given that HD~195720 is a distant giant star (spectral type M2--3~III from
SIMBAD), if associated, the companion would be a main-sequence K star.
Because of the large magnitude
difference, the presence of the projected companion does 
not affect our PSF subtraction or the analysis of the AU~Mic circumstellar disk.  

\section{DUST DISK MODELING \label{sec_model}}

It has already been suggested that the structures in the AU~Mic disk
(dust concentrations, gaps, vertically displaced clumps) are likely 
signposts of the existence of perturbing 
planetary-mass bodies in the AU~Mic
disk \citep{liu04}.  The proposed clearing in the disk inwards of 17~AU
\citep{liu_etal04} supports such a hypothesis.  From our imaging data we
cannot trace the disk to separations $<$17~AU to directly test 
the existence of a gap.  However, by combining
our knowledge of the optical to sub-mm data on the AU~Mic debris disk with an
appropriate model, we can still probe some of the physical
properties of the disk, including the size of the inner gap.  In this 
section we present results from 
a three-dimensional continuum radiative transfer code, MC3D 
\citep{wolf_henning00, wolf03}, to simultaneously model the AU~Mic SED and 
the scattered light from the disk, and to place constraints on the dust 
grain size distribution, the radial particle density distribution, and the 
inner disk radius.

\subsection{Model and Method \label{sec_method}}

The MC3D code is based on the Monte Carlo method and solves the
radiative transfer problem self-consistently.  It estimates the spatial
temperature distribution of circumstellar dust, and takes into account
absorption and multiple scattering events.  Given the non-vanishing
mid-plane optical depth of the AU~Mic disk (Section~\ref{sec_tau}), we 
believe that the use of a multi-scattering approach is warranted.
The code employs the
concept of enforced scattering \citep{cashwell_everett59}, where in a medium
of optical depth $\tau$, a fraction $e^{-\tau}$ of each photon
leaves the model space without interaction, while the remaining part
($1-e^{-\tau}$) is scattered.  The code is therefore applicable to the
low-density environments of circumstellar debris disks.  The dust
grains are assumed to be spherical with a power-law size distribution,
$n(a)\propto a^{-3.5}$ \citep*{mathis_etal77}.  We used a
standard inter-stellar medium (ISM) mixture of 62.5\% astronomical silicate 
and $25\%+12.5\%$ 
graphite \citep{draine_malhotra93, weingartner_draine01}, with
optical properties from \citet{draine_lee84}.
The extinction and scattering cross sections, and the scattering
distribution function are modeled following the Mie scattering algorithm
of \citet{bohren_huffman83}.

We use the MC3D code to model the $R$ and $H$ band scattered light 
in the AU~Mic disk, and the IR to sub-mm excess in the
SED.  The AU~Mic photosphere is best approximated by a
3600~K NextGen model \citep*{hauschildt_etal99}, as expected from
its spectral type (M1~V).  The fit was performed over the 1--12$\micron$
wavelength range, where the emission is photospheric.  Data from the
literature at shorter wavelengths were ignored, as they are not taken 
simultaneously, and hence are
strongly affected by the $V=0.35$~mag variability of the star.
By matching the model $K_S$-band flux density
to that of a blackbody of the same temperature, and adopting the {\sl
Hipparcos} distance of 9.94~pc to the star, we find that its luminosity and 
radius are $0.13L_\sun$ and $0.93R_\sun$, respectively.  For the debris 
disk we adopt a flat (unflared) geometry with a number density profile 
proportional 
to $r^{-\gamma}$, where $r$ denotes radial distance from the star, and
$\gamma$ is a constant.  We set the
outer radius of the model to 1000~AU, so that it is larger than the size
of the $R$-band scattered light emission (210~AU), and of the JCMT/SCUBA
beam used for the sub-mm measurements (FWHM of $14\arcsec = 140$~AU).
The disk inclination and opening angle were already constrained in
Section~\ref{sec_tau}.  For our modeling
purposes we assume $i=89\degr$, $\delta=0\degr$, and a flat disk model with 
a constant scale height $h$=0.8~AU.  We find that models based on these
parameters approximate the mean observed disk thickness well.

The remaining free parameters in the disk model are the exponent of the
volume density profile $\gamma$, the dust mass $M_{\rm dust}$, the minimum 
and maximum dust grain sizes $a_{\rm min}$ and $a_{\rm max}$, and the
inner radius $r_{\rm in}$.
A fit to the mean NW and SE SBP between 17--60~AU results
in a best-fit power-law index of $-\nu=-2.3\pm0.2$, indicating that the
number density profile varies as $r^{-1.3\pm0.2}$, i.e.,
$\gamma = \nu-1 = 1.3\pm0.2$ \citep*[as is true for an edge-on disk of 
isotropically scattering grains;][]{nakano90, 
backman_etal92}\footnote{Forward scattering, to the extend to which it
is characteristic of the dust grains in the AU~Mic disk, tends to increase 
$\gamma$.  Even though forward scattering is ignored in the approximation 
$\gamma = \nu-1$, it is modeled by the MC3D code, where its
amount is determined by the input grain parameters and Mie theory.}.
This value is in agreement with the range inferred for P-R
drag dominated disks \citep[1.0--1.3; e.g.,][]{briggs62}.  Given the error on
our fit, we decide to fix the value of the power-law index at the
theoretically expected value of $\gamma=1.0$ for a continuously replenished 
dust cloud in equilibrium under P-R drag \citep*{leinert_etal83, 
backman_gillett87}.  The effects of varying $\gamma$
are considered at the end of Section~\ref{sec_degen}.

\subsection{Breaking Degeneracies in the Model Parameters \label{sec_degen}}

We subsequently follow a trial-and-error by-eye optimization scheme to
determine the values of $M_{\rm dust}$, $a_{\rm min}$, $a_{\rm max}$, and 
$r_{\rm in}$.  
With a sophisticated dust disk model containing many parameter choices, it is
possible to find combinations of parameters that have degenerate effects on the SED
and/or on the SBP.  By fitting simultaneously the SED and the 
imaging data we can avoid some, but not all, of the complications.  Here we
discuss the specific degeneracies and how we can break them via the
observational constraints in hand.  We first consider the interaction
between $M_{\rm dust}$, $a_{\rm min}$, and $a_{\rm max}$, which are
strongly degenerate.  We then consider the effect of changing $r_{\rm
in}$, which is more weakly coupled with the rest of the parameters.
Finally, we extend our discussion to consider variations in the 
power-law index $\gamma$, which is otherwise kept fixed during the
modeling.

The dust mass $M_{\rm dust}$, and the minimum and maximum grain sizes, 
$a_{\rm min}$ and $a_{\rm max}$, have degenerate effects on both the SBP 
and the SED.  Decreasing $M_{\rm dust}$, or increasing $a_{\rm min}$ or 
$a_{\rm max}$, result in a decrease in the amount of mass residing in 
small grains (the bulk of the scatterers), and lowers the flux of the SBP.  
Each of these changes similarly lowers the thermal SED flux.  However, 
dust mass variations can be disentangled from grain size variations because 
of the different magnitudes of their effects on the SBP and on the SED.  
Optically thin thermal emission is a more accurate proxy of dust mass,
whereas optical-IR scattering is more sensitive to small differences in 
the mean grain size.  We therefore constrain $M_{\rm dust}$ from
the sub-mm data, while we use the color and absolute flux of the scattered 
light to determine $a_{\rm min}$ and $a_{\rm max}$.  Here we should note
that $M_{\rm dust}$ represents only the dust mass contained in grains
comparable in size or smaller than the maximum wavelength 
($\lambda_{\rm max}$) at which
thermal emission is observed.  In the case of AU~Mic, the currently existing
longest-wavelength observations are at 850$\micron$ \citep{liu_etal04}.
Consequently, we are free to adjust $a_{\rm min}$ and $a_{\rm max}$, as
long as $a_{\rm min} \leq a_{\rm max} \lesssim 1$~mm.  We will
not consider cases for which $a_{\rm max} > 1$~mm.

As a first step in finding the optimum model parameters, we confirm the 
\citet{liu_etal04} estimate of the dust mass, $M_{\rm dust}=0.011M_\earth$, 
calculated from the 850$\micron$ flux.
This value matches the sub-mm data points for a wide range (an order of
magnitude) of minimum and maximum grain sizes, whereas changing $M_{\rm
dust}$ by a factor of $>$1.5 introduces significant discrepancies from 
the observed 850$\micron$ emission.  As a next step, we constrain 
$a_{\rm min}$ by modeling the optical-near-IR color of the scattered light in
the overlap region (50--60~AU) between the \citeauthor{kalas_etal04} 
$R$-band and our $H$-band data.  Because the size of the smallest grains 
is likely comparable to the central wavelengths of the $R$ and $H$ bands, 
$R-H$ is a sensitive diagnostic for $a_{\rm min}$.  We smooth the $R$-
and $H$-band model images to the respective image resolutions (1.1$\arcsec$ 
at $R$ and 0.04$\arcsec$ at $H$), and use the appropriate aperture widths
(1.2$\arcsec$ at $R$ [\citealt{kalas_etal04}] and 0.48$\arcsec$ [this
work]).
For 
the adopted grain size distribution we find that models with
$a_{\rm min}\approx0.5\micron$, with a probable range of 0.3--1.0$\micron$, 
best approximate the disk color.  
Having constrained $M_{\rm dust}$ and $a_{\rm min}$, we find that 
$a_{\rm max} = 300\micron$ matches best the $R$- and
$H$-band flux levels of the disk.  The probable
range for $a_{\rm max}$ is 100--1000$\micron$.  Since we
have not considered models with $a_{\rm max}>1$~mm, we cannot put an
upper limit to the maximum grain size.

\begin{figure}
\plotone{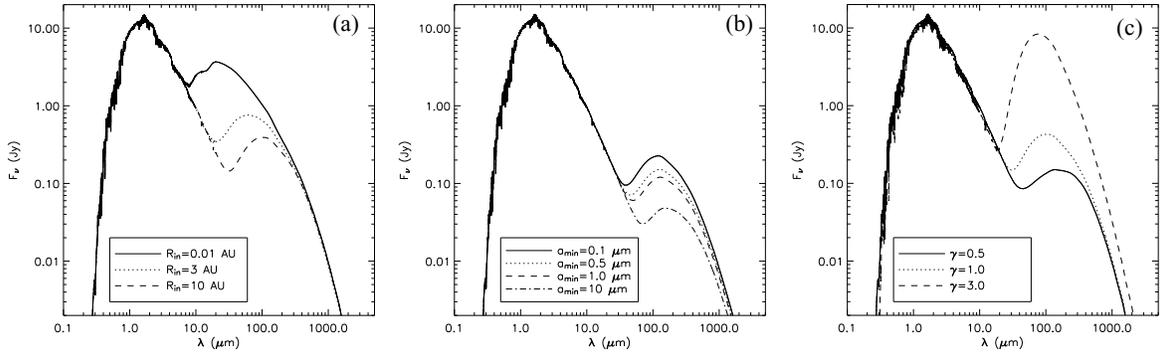}
\figcaption{Examples of degeneracies in the SED.  Larger inner gap radii (a)
are degenerate with larger minimum grain sizes (b) and shallower number
density distributions (c).  Various combinations of these parameters can
produce roughly the same SED.  The normalizations of the emitted mid-IR flux from
the disk are different among the three panels.
\label{fig_degen}}
\end{figure}

The inner radius $r_{\rm in}$ of the disk is degenerate with the mean
dust grain size in the SED.  Greater values of $r_{\rm in}$ decrease the 
mid-IR flux and shift the peak of the excess to longer wavelengths
(Figure~\ref{fig_degen}a), as do greater values of $a_{\rm min}$
(Figure~\ref{fig_degen}b) and $a_{\rm max}$.  Because the AU~Mic disk is
optically thin in the mid-plane (Section~\ref{sec_tau}), the inner disk 
radius has no effect on the flux and color of the SBP, facilitating the
isolation of the $r_{\rm in}$ parameter.  Having already determined
$a_{\rm min}$ and $a_{\rm max}$, we find $r_{\rm in}\approx10$~AU:
smaller than the 17~AU gap estimated from the single-temperature
blackbody fit in \citet{liu_etal04}.
A firm upper limit of $r_{\rm in}<17$~AU can be set based on the fact 
that we do not observe a decrease in the intensity of the scattered light with 
decreasing separation down to 17~AU \citep[Figure~\ref{fig_sbps}; 
see also Figure~2 in][]{liu04}.  

Finally, although we had fixed the value of the power-law index
$\gamma$, a brief discussion of 
its variation is warranted given the change in the SBP with radius.
The action of $\gamma$ on the SED is degenerate with the dust size, and 
with the radius of the inner gap.  Larger values of $\gamma$ are
degenerate with smaller particles and smaller inner gap radii
(Figure~\ref{fig_degen}).  Given that at $\leq$33~AU the power-law 
index of the SBP decreases to $-1.2$ (i.e., $\gamma\approx0.2$), and
that most of the mid-IR flux is produced close (10--20~AU) to AU~Mic, the 
inner disk clearing may therefore be smaller than 10~AU in radius.  Indeed,
recent {\sl HST} scattered light imaging \citep{krist_etal04} detects
the disk in to 7.5~AU (although the authors invoke forward scattering to
account for the apparent filling in of the inner gap).  Compounding
this with evidence for an increasing minimum grain size with decreasing
radial separation (Section~\ref{sec_grains}), we find that inner gap
sizes as small as $r_{\rm in} \sim 1$~AU cannot be ruled out.

\begin{figure}
\plotone{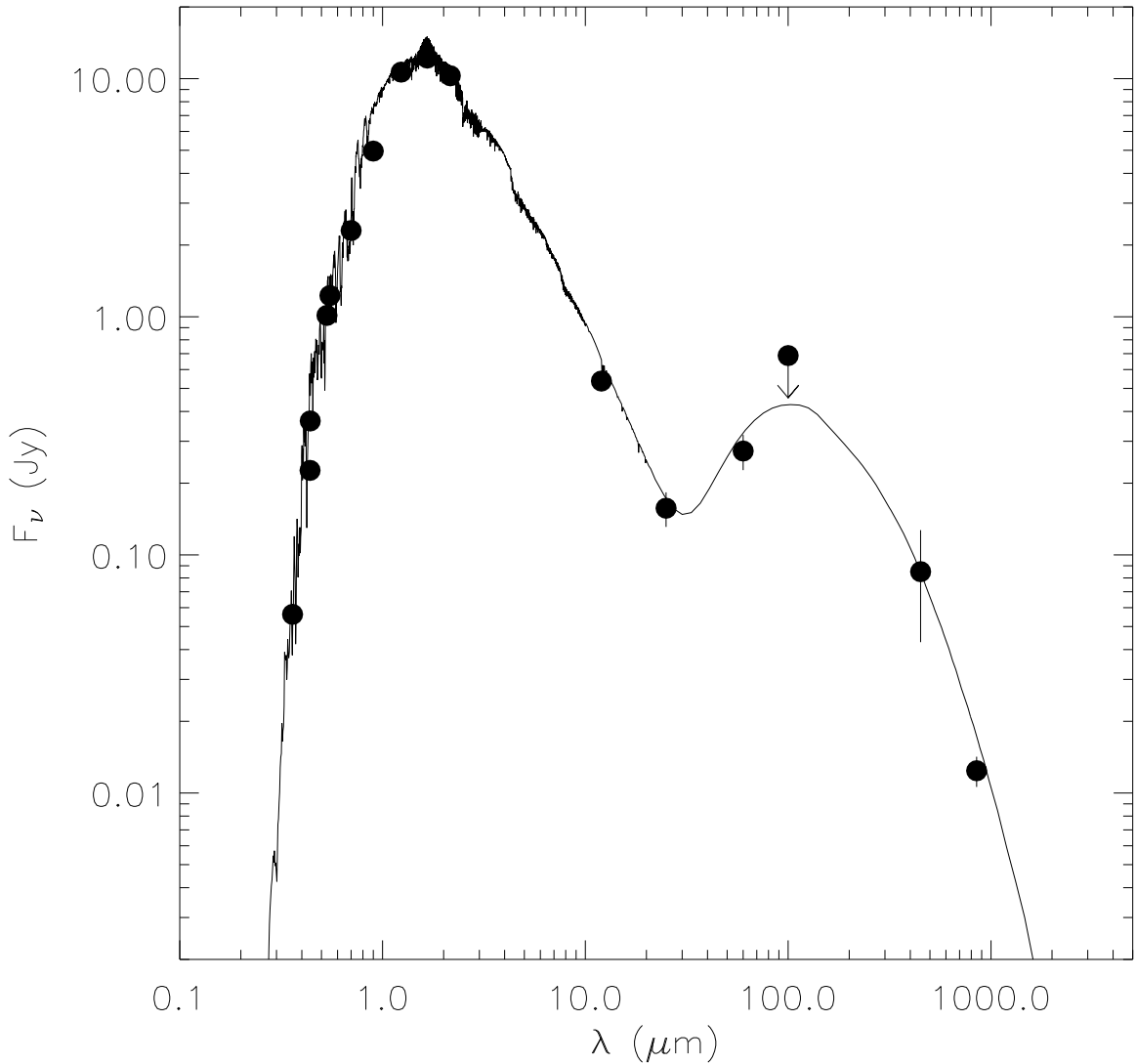}
\figcaption{SED of AU~Mic \citep[data points from][and references 
therein]{liu_etal04}.  The photophere is fit by a 3600~K NextGen model
\citep{hauschildt_etal99}, and the circumstellar excess emission is fit 
using the MC3D code (Section~4)
with model parameters listed in
Table~\ref{tab_model}.
\label{fig_sed}}
\end{figure}

Table~\ref{tab_model} lists our preferred model parameters for the AU~Mic
star-disk system.  The optical depth of the model along the disk mid-plane 
is $\tau_\parallel \approx 0.08$ at both $R$ and $H$ bands,
in agreement with the estimates in Section~\ref{sec_tau}. 
The model SBP and SED are over-plotted on the data in Figures~\ref{fig_sbps}
and \ref{fig_sed}, respectively.  
Figure~\ref{fig_contours}c shows the noiseless scattered light model of 
the AU~Mic disk with the same greyscale and contour
spacing as the image in Figure~\ref{fig_contours}a.  The model disk extends to
larger radial separations than the AU~Mic disk: an effect of the steeper power
law of the AU~Mic SBP at $>$33~AU.

\section{DISCUSSION}

Two important new results are evident from our simultaneous modeling of the
SBP and SED of the AU~Mic debris disk: (1) there is a pronounced lack of small
($<$0.3$\micron$) grains in the inner disk, and (2) the radius of 
the inner clearing may be smaller (1--10~AU) than estimated (17~AU) from 
a simple black-body fit to the IR excess \citep{liu_etal04}.  The latter 
point was already discussed
in Section~\ref{sec_degen}, and, given the shortcomings of our model in
reproducing the changing slope of the SBP, will not be belabored further.
Here we discuss the derived minimum grain size along with recent evidence 
for its
dependence on disk radius.  We then focus on the change in slope of the
SBP of the AU~Mic debris disk, and draw a parallel with the $\beta$~Pic
system.  We propose that identical dynamical processes in the two debris 
disks can explain the observed homology.

\subsection{Minimum Grain Size as a Function of Disk Radius \label{sec_grains}}

From the model fits to the color and absolute flux of the scattered light
from the AU~Mic debris disk, we find that the dust grains are between
$a_{\rm min} = 0.5_{-0.2}^{+0.5} \micron$ and $a_{\rm
max}=300_{-200}^{+700} \micron$ in size (although the 1~mm upper limit on
the maximum grain size is not robust).
In reality, our constraints on the grain parameters are valid only over 
the 50--60~AU region, where we have information from both the $R-H$ color of 
the scattered light and the SED.  We
have very few constraints for the outer disk ($>60$~AU), which is seen 
only in $R$-band scattered light and is too cold to be detected in emission
at wavelengths $<$1~mm.

Shortly before receiving the referee report for this paper, sensitive
high-resolution 0.4--0.8$\micron$ images of AU~Mic became available from {\sl
HST} \citep{krist_etal04}.  These show the debris disk over 7.5--150~AU 
separations from the star and thus provide complete overlap with 
our AO data.  A brief discussion of the two data sets in the context of the
minimum grain size is therefore warranted.  For consistency with
\citeauthor{krist_etal04},
we re-did our surface photometry with the $0.25\arcsec\times0.25\arcsec$ 
apertures used by these authors.  Because the two data sets have similar
angular resolution (0.04$\arcsec$ vs.\
0.07$\arcsec$), and because the aperture size is much larger than the
FWHM of the PSFs, the systematics of the photometry should be negligible.
The \citeauthor{krist_etal04} {\sl HST} $F606W$ (0.59$\micron$) data are 
consistent with the
\citeauthor{kalas_etal04} ground-based $R$-band (0.65$\micron$) observations 
over 50--60~AU from the
star, and thus our conclusions about the grain sizes in this region
remain unchanged.  However, a comparison of the $H$-band and the
$F606W$-band SBP over the region of overlap shows that the $F606W-H$ color
changes from 2.9$\pm$0.2~mag (i.e., approximately neutral, since $R-H=2.9$ 
for AU~Mic) at 17--20~AU to 2.0$\pm$0.3~mag at 50--60~AU.  That is, the 
debris disk becomes increasingly bluer at larger radii.  The effect is 
gradual and is also reported in
\citeauthor{krist_etal04}, where it is observed over a narrower wavelength
range (0.4--0.8$\micron$) at 30--60~AU from the star.  

The neutral color of the dust at 20~AU indicates that the majority of 
scatterers there are larger than 1.6$\micron$.  Compared with the minimum 
grain size ($0.5_{-0.2}^{+0.5}\micron$) that we derived at 50--60~AU,
this indicates that at smaller separations grains are bigger.
Such a dependence of grain size on radius would further imply that
the radius at which the SBP changes power-law index should be
wavelength-dependent, occurring farther away from the star at shorter
wavelengths.  Evidence for this may indeed
be inferred from a comparison between the {\sl HST} and Keck AO data: in
the $F606W$ ACS filter the break in the SBP is seen at $\approx$43~AU 
\citep{krist_etal04}, whereas at $H$-band it occurs near 33--35~AU 
\citep[Section~\ref{sec_morph};][]{liu04}.  The indication
that the minimum grain size decreases with disk radius is thus confirmed
from two independent observations.

Particles smaller than $a_{\rm min}$ may be removed as a result of either 
coagulation into larger particles (grain growth), destruction by P-R 
and/or corpuscular drag, or radiation pressure blow-out.  Given that 
$a_{\rm min}$ is larger than the radiation pressure blow-out size 
(0.14$\micron$, for a radiation pressure to gravity ratio $\beta$=0.5 
and a grain density of 2.5~g~cm$^{-3}$), grain collisions and drag 
forces dominate 
the dynamics of $>$0.14$\micron$ grains around AU~Mic.  Therefore,
the origin of the sub-micron grains scattering visible light at
wide separations (where the collision and P-R time scales are longer than 
the age of the
star) may also be primordial: rather than being blown out from the inner disk,
these grains may be remnant from the proto-stellar cloud that never
coagulated beyond an ISM grain size distribution.  

\subsection{The Change in the SBP Power-Law Index: a Comparison with
$\beta$~Pic}

It is not surprising that our preferred model cannot reproduce the
detailed structure of our high angular resolution IR image.  The
model parameters were found only after a coarse sampling of the
parameter space, and through a number of
simplistic assumptions that merely approximate the physical conditions in
the AU~Mic debris disk.  In particular, under the assumption of a uniform 
grain size and density distribution over 10--1000~AU,
the MC3D model cannot mimic the SBP slope change at
$\sim$33~AU and the clumpy substructure over 17--60~AU described in 
\citet{liu04} and confirmed in Section~\ref{sec_morph}.
While the dust clumps are high-order
perturbations that may require dynamical considerations for proper modeling,
the change in the SBP potentially could be explained in 
the framework of existing dust disk scenarios.

The occurrence of the power-law break at similar
radii in the SBPs both arms of the projected disk suggests that this
is a ring-like structure surrounding the star, rather than a
discrete feature at one location in the disk.  Models involving 
dynamical interaction with planets have been proposed to explain
ring-like structures in circumstellar disks 
\citep[e.g.,][]{roques_etal94, liou_zook99, kenyon_etal99,
kenyon_bromley04},
and the clumpy structure of the AU~Mic disk does suggest the presence of
unseen planets \citep{liu04}.  However, such models tend to produce 
discrete rings, as around HR~4796A and HD~141569A, 
rather than the radially dimming SBP of the AU~Mic disk.
Similar changes in the power-law index have 
also been seen in the SBPs of other resolved circumstellar 
disks: $\beta$~Pic (\citealt*{artymowicz_etal90}, \citealt{heap_etal00}), 
TW~Hya \citep{krist_etal00, weinberger_etal02}, and HD~100546 
(\citealt*{pantin_etal00}; \citealt{augereau_etal01}).  A 
different mechanism, not necessarily involving planets, may be at play in
these systems.

The TW~Hya and HD~100546 circumstellar disks are gas-rich and have
large mid-plane optical depths, and hence are very much unlike the
gas-poor \citep{roberge_etal04}, optically thin AU~Mic debris disk.
However, a comparison with $\beta$~Pic is particularly illuminating,
because of the similar viewing geometry of the two systems, and their 
identical ages.   In the remainder of this
section we seek a common disk architecture that can self-consistently
account for the broken power-law morphology of the SBPs of these two debris
disks.

The SBP of $\beta$~Pic exhibits a break at 5--6$\arcsec$ (100--120~AU) 
from the star (e.g., \citealt*{golimowski_etal93}, \citealt{heap_etal00}), 
with similar values (from --1 to --4) of the power-law index on either side of 
the break as in the SBP of AU~Mic.  From $K\prime$-band (0.21$\micron$) AO 
observations resolving the $\beta$~Pic disk over 1.5$\arcsec$--6$\arcsec$, 
\citet{mouillet_etal97} observe the break at a somewhat smaller
radius, 4--4.5$\arcsec$ (75--85~AU), with a smaller change in the SBP 
power-law index (from --1 to --3).  This may be the inner edge of
the SBP break observed in the visible, or may indicate a 
wavelength-dependence of the $\beta$~Pic break radius similar
to the one potentially seen in the AU~Mic disk (Section~\ref{sec_grains}).
However, because the \citeauthor{mouillet_etal97} $K\prime$-band data do 
not extend beyond the optical break radius (6$\arcsec$), and because the 
change in the power-law index observed at $K\prime$ does not span the full 
range of power-law indices inferred from optical imaging, this data set 
will not be considered further.

\citet{artymowicz_etal89} model the break in optical SBP of the $\beta$~Pic 
disk using two different power laws for the 
number density of dust particles in the disk, for radii
less than or greater than 100~AU, respectively.  \citet{backman_etal92} 
consider the possibility of differing grain sizes, in addition.  While
either model may correctly describe the architecture of the $\beta$~Pic
disk, both are purely phenomenological, as they do not model the physics 
behind the discontinuity in the disk.  Based on the
apparent homology between these two debris disks, we
believe that a plausible two-component model should be able to explain
both systems self-consistently.  In light of this, several physical scenarios 
from the subsequent literature are considered below.  We find that none 
of them offer a unique explanation, and propose separate hypotheses in
Sections~\ref{sec_collisions} and \ref{sec_PR}.

\subsubsection{Ice or Comet Evaporation \label{sec_ice}}

\citet{backman_etal92} and \citet*{pantin_etal97} suggest that the
discontinuity in the SBP of $\beta$~Pic
may correspond to the location of the ``ice boundary'' in the disk:
all dust particles at separations $>$100~AU are covered 
with ice, while at shorter separations some may not be.  This results in a
deficiency of highly reflective particles in the inner regions, creating 
a shallower power-law index for the scattered light profile.  In an 
optically thin disk the radius of the ice boundary should scale 
as the square root of the stellar luminosity.  Adopting 
$L_{\rm AU Mic}=0.13 L_\sun$ (Table~\ref{tab_model}), and 
$L_{\beta {\rm Pic}}=8.7L_\sun$, and assuming identical grain chemistry
in the two systems, we find that the corresponding boundary around
AU~Mic should scale down to a radius of 13--15~AU; too close to
account for the break at 33~AU.  

An alternative hypothesis, involving dust extraction through evaporation 
of gas from a reservoir of cometary bodies around $\beta$~Pic is proposed by 
\citet*{lecavelierdesetangs_etal96}.  However, the radial distance of
this evaporating reservoir should scale in the same manner as that of the ice 
boundary.  Hence, neither of these two hypotheses can be applied
simultaneously to AU~Mic and $\beta$~Pic.

\subsubsection{A Belt of Parent Bodies \label{sec_exoab}}

A reservoir of parent bodies at a discrete range of separations from
AU~Mic could explain the kink in the SBP.
\citet{gorkavyi_etal97} calculate that the Main asteroid belt in the
Solar System should produce a break in the power-law index of the
number density distribution of interplanetary grains from $-1.3$ to $-6.4$ at 
0.5--3.0~AU from the Sun.  These predictions are consistent with empirical data 
from radar meteors and from impact detectors on spacecraft
\citep{divine93}.  The inner edge of this belt of asteroids in the AU~Mic and
$\beta$~Pic 
systems would be at the location of the kinks, at $\sim$33~AU and $\sim$110~AU
from the stars, respectively.
By continuing the analogy with the Solar System, such belts of parent
bodies would likely need to
be maintained in a discrete range of orbits through mean motion resonances
with planets \citep[e.g.,][and references therein]{liou_zook97}.  That is, this
particular scenario may provide further indication for the existence of
planets in the two disks, in addition to the evidence arising from the 
their clumpy structure.  However, this model is poorly constrained, as we 
are free to invoke a belt of parent bodies at any distance from either star.

\subsubsection{Collisional Evolution \label{sec_collisions}}

For two stars of the same age, the disk around the more massive star is
expected to be collisionally evolved out to a greater radius, because of the
inverse scaling of the orbital period (and hence, collision frequency) 
with orbital radius and stellar mass.  The
collisional time scale for particles of mean size $a$ on a circular orbit of 
radius $r$ from a star of mass $M_\ast$ is
\begin{equation}
t_{\rm coll} \sim \frac{P}{4\pi^2 a^2 r n(r)} \propto 
\frac{r^{1/2}}{M_\ast^{1/2} n_0 r^{-\gamma}} = 
\frac{r^{\gamma+1/2}}{M_\ast^{1/2} n_0},
\end{equation}
where $P$ is the orbital period, and $n_0$ is
the normalization constant for the number density distribution, that
we presume scales as $M_{\rm dust}$.  Given the approximate
ratios of the stellar masses of AU~Mic and $\beta$~Pic (0.28, assuming
$M_{\beta~{\rm Pic}}=1.8 M_\sun$), and of their circumstellar dust masses, 
($0.2\pm0.1$, where the mass of the $\beta$~Pic disk was taken as the
average of the estimates from \citet*{sheret_etal04} and \citet{dent_etal00})
and assuming 33~AU and 110~AU as the radii of the kinks in the corresponding
SBPs, we find that for $\gamma=1.5^{+0.3}_{-0.4}$ the collisional time 
scales at the respective
separations around the two stars are equal.  This would imply SBPs decreasing 
approximately as $r^{-\gamma-1} = r^{-2.5}$, which is within the
ranges found in Section~\ref{sec_morph} for AU~Mic and in \citet{heap_etal00}
for $\beta$~Pic, and agrees with the fit to the mean AU~Mic $H$-band SBP 
(power-law index of $-2.3\pm0.2$; Figure~\ref{fig_sbps}).
Therefore, in this scenario the two disk systems scale correctly within 
the errors, indicating that whatever process we are observing may scale with 
the mean time between inter-particle collisions.  Although the
collisional time-scale at the location of the break in the AU~Mic SBP is
considerably shorter (1--3~Myr) than the age of the star (8--20~Myr), 
other, slower
processes in the disk, e.g., grain growth, may scale linearly with the 
time between particle collisions.  As noted in Section~\ref{sec_grains}, 
grain growth could also explain the observed dependence of grain size
with orbital radius.

\subsubsection{Poynting-Robertson Drag \label{sec_PR}}

The observed change in the power-law index of the SBP may be a reflection
of the finite lifetimes of sub-micron grains in the inner disk.  Having ruled 
out radiation pressure as a dominant force on grains larger than 0.14$\micron$
(Section~\ref{sec_grains}), we propose a hypothesis based on drag forces
for grain removal,
in particular P-R drag.  Although corpuscular drag may dominate 
the dynamics of dust around M stars \citep*[e.g.,][]{fleming_etal95}, 
the strength of stellar winds from M dwarfs remains
largely unknown.  We therefore ignore corpuscular drag in the following
analysis \citep*[though see][for a discussion of the role of corpuscular
drag in the AU~Mic disk]{plavchan_etal05}, 
and consider only P-R drag.  As long as the magnitude of the
corpuscular drag force around AU~Mic is not much greater than the
magnitude of the P-R drag force, the conclusions remain unchanged.

If P-R drag was responsible for the depletion of micron-sized grains in 
the inner disks of AU~Mic and $\beta$~Pic, then the P-R lifetime 
($t_{\rm PR}$) of the smallest grains $a_{\rm min}$ should be constant 
as a function of disk radius $r$, and should equal the age of the stars 
($t_{\rm age}$).  This can be inferred from the expression for the P-R 
lifetime of a particle of size $a$ \citep*[see, e.g.,][]{burns_etal79}:
\begin{equation}
t_{\rm PR}(a, r) = \left (\frac{4 \pi a \rho}{3} \right ) \left (\frac{c^2
r^2}{L_*} \right ),
\end{equation}
where, $\rho$ is the mean grain density (2.5~g~cm$^{-3}$ for silicates),
and $c$ is the speed of light.  Based on the assumption that P-R drag is 
the dominant removal mechanism for grains larger than the blow-out size
(0.14$\micron$), the P-R lifetime of the smallest grains is
$t_{\rm PR}(a_{\rm min}, r) = t_{\rm age} = {\rm const}$.  Note, however, 
that the size of the smallest grains, $a_{\rm min}$, is not 
a constant, but varies as $a_{\rm min} \propto r^{-2}$.

For AU~Mic we found $a_{\rm min} \geq
0.3\micron$ at 50--60~AU, and $a_{\rm min} \geq 1.6\micron$ at 17--20~AU
(Section~\ref{sec_grains}).  We obtain $t_{\rm PR}(0.3\micron, 50{\rm
AU}) = 9.9$~Myr and $t_{\rm PR}(1.6\micron, 20{\rm AU}) = 8.4$~Myr.
We do not have information about the change in $a_{\rm min}$ as a function 
of radius in $\beta$~Pic.  We only note that from mid-IR and visual
images, \citet{artymowicz_etal89} find that they require 
``few-micron-sized'' silicate grains\footnote{A second solution involving
1--15$\micron$ ice grains is found to be equally plausible.  However, its 
grain properties differ widely from the ones adopted for the AU~Mic
circumstellar dust in this paper.} to model the scattered light at
$>$6.0$\arcsec$ ($>$115~AU) from the star.  Assuming a minimum grain size
of 3$\micron$, we find $t_{\rm PR}(3\micron, 115{\rm AU}) = 7.8$~Myr
around $\beta$~Pic.  Given the uncertainty in $a_{\rm min}$,
the obtained P-R time-scales are not constrained to better than a factor
of 1.5--2.  Nevertheless, they are remarkably similar and agree well 
with the ages of AU~Mic and $\beta$~Pic.  

\subsubsection{Summary of Proposed Scenarios}

We find that dynamical scenarios based on collisional evolution or P-R 
drag in debris disks offer simpler and more self-consistent accounts of 
the homology between the SBPs of the AU~Mic and $\beta$~Pic debris disks, 
compared to scenarios relying on ice/grain evaporation or belts of 
orbiting parent bodies.  Moreover, both our hypotheses can account for the
inferred decrease in the minimum grain size with increasing separation
from AU~Mic.  We therefore conclude that both 
are plausible.  Given the similarities in their predictions, we do not
single out which one of them is more likely, but defer that analysis to
a more detailed theoretical work.  Regardless of which of the two
processes is found to be dominant, we can confidently claim that optically
thin circumstellar disks exhibiting breaks in their SBPs are observed 
in transition between a primordial and a debris state.

\section{CONCLUSION}

We have used AO $H$-band observations of scattered light to probe the
morphology of the debris disk at 17--60~AU from the young nearby 
M dwarf AU~Mic.  We find that the disk is within $\sim$1$\degr$ of edge-on,
and that it exhibits a number of morphological peculiarities: radial
asymmetry, spatially resolved clumps, and a change in
the power-law index of the surface brightness profile near 33~AU.  The
observed morphology agrees with that reported in \citet{liu04}, and is
suggestive of the existence of planetary perturbers in the disk.
No planets are detected down to $1 M_{\rm Jup}$ at $>$20~AU.  

We use a Monte Carlo three-dimensional dust disk model to constrain the 
overall disk parameters, by optimizing them against the AU~Mic SED, 
and near-IR and optical scattered light images of the disk \citep[this
paper;][]{kalas_etal04}.  The combined use of SED data tracing the thermal
emission from large grains, and of imaging data tracing grain properties,
allows us to break several important degeneracies in models of
circumstellar disks that cannot be resolved using only one of the two data
sets.  From the SED we confirm the previously inferred \citep{liu_etal04}
circumstellar dust mass of 0.011$M_\earth$, and from the properties 
of the scattered light we infer that the debris particles are 
$\geq0.5_{-0.2}^{+0.5}\micron$ in size at 50--60~AU from
AU~Mic.  We find tentative evidence for a maximum grain size of
$300_{-200}^{+500}\micron$.  However, since the data are not sensitive to 
particles $\gg$1~mm, the result is consistent with no upper limit on 
the grain
size.  Assuming a single dust size and density distribution, we estimate 
that the radius of the inner disk clearing is 10~AU.  However,
smaller ($\sim$1~AU) gap sizes cannot be ruled out if a shallower density 
profile (as observed inwards of 33~AU) and/or larger grains in the inner 
disk (as evidenced from the optical-near-IR color of the disk) are
adopted.
We attribute the lack of sub-micron particles in the inner disk either to 
grain growth, or destruction by P-R and/or corpuscular drag (for grains
$>$0.14$\micron$), or to blow-out by radiation
pressure (for grains $<$0.14$\micron$). 
All of these mechanisms can explain the increase in 
relative density of small grains with increasing radius in the disk.

The MC3D model can account for the overall disk profile and colors to
first order,
although our one-component model fails to reproduce higher-order
effects, such as the change in power-law index of the SBP.  We have discussed
a number of scenarios which may be capable of reproducing such a change
in the joint context of the AU~Mic and $\beta$~Pic debris disks.
We find that
models dividing the disk into two separate components, with different grain
distribution and/or composition, represent the combined SED and imaging data
best.  In particular, classes of models which scale with the
collision or the P-R time-scale are most likely to explain both debris
disks self-consistently.  

Future high-dynamic range imaging observations probing closer to AU~Mic
(e.g., with nulling interferometry in the mid-IR) will further narrow down 
the architecture of its debris disk.  Photometric and spectroscopic 
observations with {\sl Spitzer} could better constrain the SED of AU~Mic, 
and could be used to look for spectroscopic
features.  These could trace small amounts of dust and gas in the inner disk,
even if no continuum excess is seen at $<25\micron$.

\acknowledgements

We are grateful to Randy Campbell, Paola Amico and David
Le Mignant for their guidance in using Keck AO, Keith Matthews and Dave
Thompson for advice on using NIRC2, and our telescope operator Madeline
Reed at the Keck~II telescope.  We also thank the anonymous referee for
his/her constructive comments, Scott Kenyon for a critical review of
the draft manuscript, and Laird Close and Peter Plavchan for insightful
discussions.  This publication has made use of data 
products from the Two Micron All Sky
Survey, which is a joint project of the University of Massachusetts and
the IPAC/California Institute of Technology, funded by the NASA and the
NSF, and of the SIMBAD database, operated at CDS, Strasbourg, France.
Finally, the authors wish to extend special
thanks to those of Hawaiian ancestry on whose sacred mountain of Mauna
Kea we are privileged to be guests.  Without their generous hospitality,
none of the observations presented herein would not have been possible.
J.A.E.\ acknowledges support from a Michelson Graduate Research Fellowship.
S.W.\ was supported by the German Research Foundation (DFG) through the Emmy
Noether grant WO~857/2--1.


\clearpage

\clearpage

\clearpage

\begin{deluxetable}{lcc}
\tablewidth{0pt}
\tablecaption{Preferred model parameters for the AU~Mic system. \label{tab_model}}
\tablehead{\colhead{Parameter} & \colhead{Value} & \colhead{Range}}
\startdata
Stellar luminosity $L_\ast$ & 0.13$L_\sun$ & fixed \\
Stellar radius $R_\ast$ & 0.93$R_\sun$ & fixed \\
Stellar temperature $T_{\rm eff}$ & 3600~K & fixed \\
Dust mass $M_{\rm dust}$ & 0.011$M_\earth$ & 0.008--0.016$M_\earth$ \\
Dust size distribution & $a^{-3.5}$ & fixed \\
Number density profile & $r^{-1.0}$ & $-0.2$ to $-3.0$, 
	fixed\tablenotemark{\dag} \\
Outer disk radius $r_{\rm out}$ & 1000~AU & fixed \\
Inner disk radius $r_{\rm in}$ & 10~AU & 1--10~AU \\
Inclination angle $i$ & 89$\degr$ & $\gtrsim89\degr$ \\
Scale height $H$ & 0.8~AU at $r=40$~AU & $\lesssim$1.0~AU at $r=40$~AU\\
Opening angle $2\delta$ & 0$\degr$ & $\lesssim6\degr$ \\
Minimum grain size $a_{\rm min}$ & 0.5$\micron$ & 0.3--1$\micron$ \\
Maximum grain size $a_{\rm max}$ & 300$\micron$ &
100--1000$\micron$\tablenotemark{\ddag}
\enddata
\tablenotetext{\dag}{Treated as a fixed parameter during the MC3D model
fitting.  The listed range corresponds to the range of fits to
the SBP over 17--60~AU.}
\tablenotetext{\ddag}{The SED data are not sensitive to emission from
grains $\gtrsim$1000$\micron$ in size, so we have not run models with
$a_{\rm max}>1000\micron$.}
\end{deluxetable}

\end{document}